\documentclass[%
reprint,
superscriptaddress,
 amsmath,
 amssymb,
pra,
floatfix,
]{revtex4-1}

\usepackage{multirow}
 \usepackage{rotating} 
\usepackage{hyperref}
\usepackage{booktabs}
\usepackage{adjustbox}
\usepackage{amsmath}
\usepackage{optidef}
\usepackage{qcircuit}
\usepackage{graphicx}%
\usepackage{subfigure}
\usepackage{dcolumn}%
\usepackage{bm}%
\usepackage{dsfont}
\usepackage{xcolor}
\usepackage{placeins}
\usepackage{tikz}
\definecolor{C0}{HTML}{1F77B4}
\definecolor{C1}{HTML}{FF7F0E}
\definecolor{C2}{HTML}{2CA02C}
\definecolor{C3}{HTML}{D62728}
\definecolor{C4}{HTML}{9467BD}
\definecolor{C5}{HTML}{8C564B}
\definecolor{C6}{HTML}{E377C2}
\definecolor{C7}{HTML}{7F7F7f}
\definecolor{C8}{HTML}{BCBD22}
\definecolor{C9}{HTML}{17BECF}

\begin{document}

\title{Quantum Machine Learning Framework for Virtual Screening in Drug Discovery: a Prospective Quantum Advantage}

\author{Stefano Mensa}
\thanks{These authors contributed equally.}
\affiliation{The Hartree Centre, STFC, Sci-Tech Daresbury, Warrington, WA4 4AD, United Kingdom}

\author{Emre Sahin}
\thanks{These authors contributed equally.}
\affiliation{The Hartree Centre, STFC, Sci-Tech Daresbury, Warrington, WA4 4AD, United Kingdom}

\author{Francesco Tacchino}
  \affiliation{IBM Quantum, IBM Research -- Zurich, S\"aumerstrasse 4, R\"uschlikon, CH–8803, Switzerland}
  
\author{Panagiotis Kl. Barkoutsos}
  \affiliation{IBM Quantum, IBM Research -- Zurich, S\"aumerstrasse 4, R\"uschlikon, CH–8803, Switzerland}
  
\author{Ivano Tavernelli}
\email{ita@zurich.ibm.com}
  \affiliation{IBM Quantum, IBM Research -- Zurich, S\"aumerstrasse 4, R\"uschlikon, CH–8803, Switzerland}


\date{\today}

\begin{abstract}

Machine Learning (ML) for Ligand Based Virtual Screening (LB-VS) is an important \textit{in-silico} tool for discovering new drugs in a faster and cost-effective manner, especially for emerging diseases such as COVID-19. In this paper, we propose a general-purpose framework combining a classical Support Vector Classifier (SVC) algorithm with quantum kernel estimation for LB-VS on real-world databases, and we argue in favor of its prospective quantum advantage. Indeed, we heuristically prove that our quantum integrated workflow can, at least in some relevant instances, provide a tangible advantage compared to state-of-art classical algorithms operating on the same datasets, showing strong dependence on target and features selection method. Finally, we test our algorithm on IBM Quantum processors using ADRB2 and COVID-19 datasets, showing that hardware simulations provide results in line with the predicted performances and can surpass classical equivalents.

\end{abstract}

\maketitle

\section{Introduction}\label{sec1}

The early stages in a drug discovery process are notoriously expensive and time-consuming, usually coupling wet lab with \textit{in-silico} technologies.
Virtual Screening (VS) is an important computational technique used to provide a quick and economical method for the discovery of novel therapeutics. 
In practice, VS searches digital libraries of small molecules to identify compounds that are most likely to bind to a target (\textit{e.g.}, a protein) and therefore exert a pharmacological effect (i.e.~activity). In an ideal scenario, the pharmacological binding site is known, and the structure of the target is well characterised. 
Therefore, the VS could -- in principle -- be performed directly using a digital representation of the binding site.

\begin{figure*}
\centering
\includegraphics[width=1\textwidth]{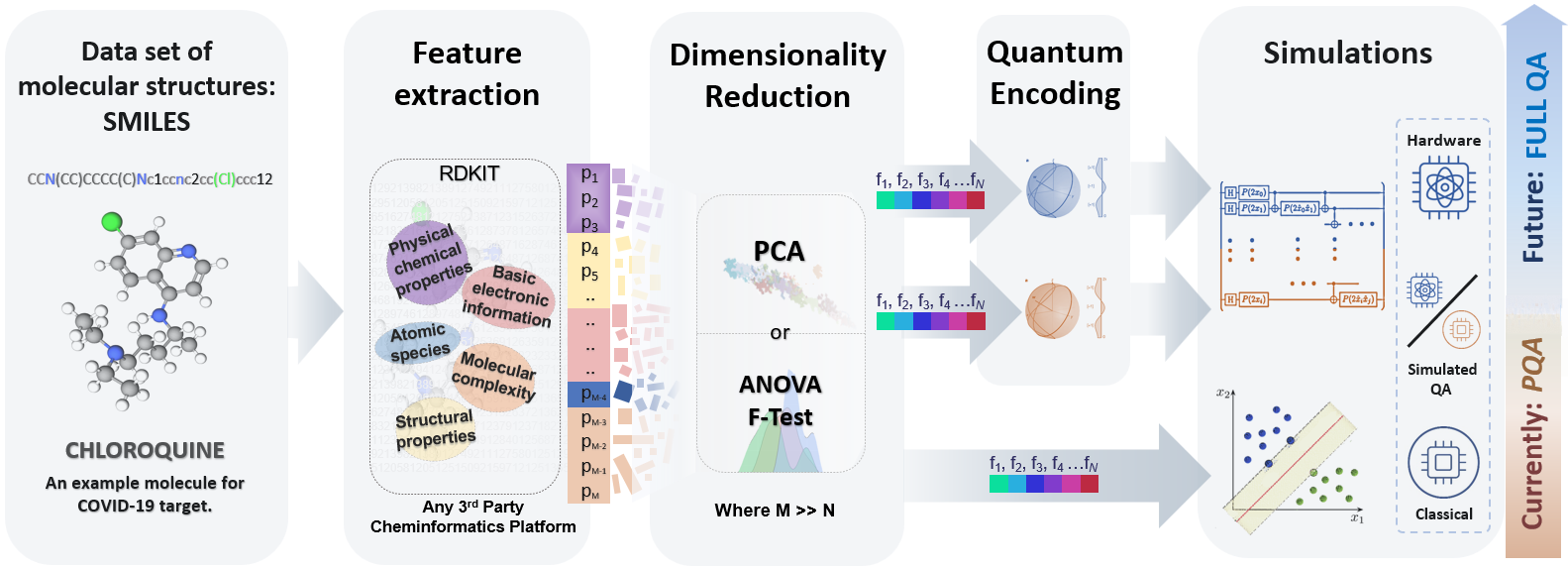}
\caption{Hybrid classical-quantum integrated Machine Learning framework for Ligand-Based Virtual Screening in drug discovery. A database of SMILES encoded molecules is used to extract a variety of molecular features using RDKit. Through different feature reduction and selection methods, we refine the feature vectors, which are then passed on to train and test a Support Vector Classifier algorithm. We compare the performance of the SVC algorithm trained using a classical and a quantum kernel on both classical and quantum hardware, and we find cases in which the simulation of quantum algorithms/hardware running  on a reduced number of  features outperforms classical equivalents. We name this \textit{Prospective algorithmic Quantum Advantage} (PQA) in drug discovery. With the advent of larger and better performing near-term quantum computers, PQA may lead to full quantum advantage on relevant problem instances.}
\label{fig1}
\end{figure*}

However, the structure of the target is generally unknown, especially for emerging diseases such as COVID-19, caused by SARS-CoV-2, the coronavirus that arose in December 2019.
In such cases, the information available for known active molecules (ligands) against a given target is commonly used to drive the design process.
Starting from the chemical structure of known ligands, it is possible to infer hundreds of cheminformatic features that can be coupled with experimental evidence, when available, and then used to perform VS of a digital database. 
Such methods are known as Ligand Based VS (LB-VS).
A conventional LB-VS strategy consists in using machine learning (ML) algorithms to train a classifier that can identify potential drug candidates from a digitalised library of compounds. The latter are usually characterized by means of a combination of theoretical and experimental features.
Thanks to their proven efficiency in identifying patterns in unstructured data, ML algorithms such as Support Vector Classifiers (SVC)  became an effective tool for performing LB-VS tasks~\cite{Jorissen2005VirtualMachine,Hert2006NewSearching,Plewczynski2009PerformanceScreening,Sato2012ApplicationInhibitors,Lavecchia2015Machine-learningApplications,Korotcov2017ComparisonSets,Lo2018MachineDiscovery,Yang2019ConceptsDiscovery,Walters2020NewScreening,Patel2020MachineDiscovery}. 

Recent progress in the field of quantum machine learning (QML)~\cite{Biamonte2017QuantumLearning,Mangini_2021} %
identified Quantum Kernel (QK) methods~\cite{Schuld2019QuantumSpaces}, and specifically Quantum Support Vector Classifiers (QSVC)~\cite{Havlicek2019SupervisedSpaces}, among the most promising candidates for extending the reach of artificial intelligence beyond classical boundaries. By mapping classical inputs into high-dimensional complex Hilbert spaces, QK methods may efficiently produce atypical patterns, thus potentially leading to quantum advantage in training speed, prediction accuracy and classification~\cite{Wu2021ApplicationLHC,Liu2021ALearning}. While many key challenges~\cite{Huang2021PowerLearning,kubler2021inductive}, and particularly the quest for robust quantum advantage on naturally occurring data sets, remain open, QML protocols appear to be very well suited in situations dealing with complex data and large amounts of available information, such as the LB-VS case~\cite{Cao2018PotentialDiscovery}.

At the same time, the application of quantum machine learning to VS comes with complications, and guidance on how to port the LB-VS problem to a quantum computing framework is currently missing. 
The first practical issue consists in identifying the most suited approach to efficiently load data  (\textit{i.e.}~the database of molecules with features) into a quantum register.  
Second, there is the need to optimize the available QML algorithms so that to exploit -- in the most efficient way -- the potential of the quantum processor for the specific LB-VS task, providing a path for achieving quantum advantage over classical counterparts.
In this paper, we propose a general framework for the integration of quantum computing in the LB-VS workflow for drug discovery, see Fig.~\ref{fig1}. We demonstrate that the construction of a quantum classifier trained with a small number of cheminformatics descriptors can, in some cases, significantly outperform classical state-of-the-art ML and Deep Learning (DL) algorithms, when working using a specific benchmarking dataset for ML/DL-based VS. 
The performance of our quantum-powered LB-VS methodology is validated on superconducting quantum processors, %
using an ADRB2 benchmarking dataset as well as a novel dataset containing COVID-19 inhibitors~\cite{Gawriljuk2021MachineSARS-CoV-2} and
reporting results in line with the numerical predictions obtained with classical numerical simulations. %
The proposed approach is not tied to a specific use case and can be applied to screen any digital database of active/inactive compounds against a desired target,  %
using any third-party cheminformatics package.

Inspired by these promising results and confident in the scalability of the quantum approach, we also introduce the new concept of ``Prospective Quantum Advantage'' (PQA) obtained when a quantum algorithm simulated on a classical computer or executed on state-of-the-art hardware (or any combination of the two) can provide -- \textit{at least in some relevant instances} -- a tangible advantage compared to the best known classical algorithm operating with the same data sets (input information).  
In addition, PQA implies no evident restrictions on its extension to larger problem sizes, for which the quantum algorithms cannot be efficiently simulated on classical computers, while keeping the same -- or even extend -- the observed advantage. When demonstrations on larger quantum processors will become feasible, PQA will lead to a full effective Quantum Advantage. 

The demonstration of PQA for a drug discovery workflow based on LB-VS is one of the main outcomes of the results presented in this work. 

\section{Methodology}\label{methodology}

We design a target-agnostic computational workflow to provide heuristic algorithmic evidence of quantum advantage over classical methods in LB-VS. 
More specifically, we use classical data to perform VS using a quantum Support Vector Classifier algorithm, hence evaluating its performance with respect to its classical counterparts.
A like-for-like comparison between classical and quantum methods is fundamental to be as fair as possible in assessing the actual potential for quantum advantage.

\subsection{Dataset}\label{Dataset}

The general nature and the purpose of our classification workflow both require a realistic benchmarking dataset specifically dedicated to applied machine learning in virtual screening.
Specifically, the dataset should~\cite{Sieg2019InScreening,Tran-Nguyen2020LIT-PCBA:Screening,Rohrer2009MaximumData,Xia2015BenchmarkingScreening} (1) imitate real-world screening libraries and guide ML methods to discriminate active from inactive compounds; (2) contain molecules with unambiguous experimentally measured activity against a target, i.e.~with either potent/moderate activity or inactivity; (3) have a realistic ratio between active and inactive molecules; (4) contain active and inactive compounds with comparable molecular properties; (5) be chemically unbiased.

The LIT-PCBA dataset \cite{Tran-Nguyen2020LIT-PCBA:Screening} encompasses all the previous points for fifteen well-characterised biological targets, and it is a natural choice for this project. 
Furthermore, and with respect to prior efforts \cite{Huang2006BenchmarkingDocking,Chaput2016BenchmarkPerformance,Wallach2018MostGeneralization,Sieg2019InScreening}, the LIT-PCBA dataset was created specifically to test the performance of ML tasks in VS of molecules. The dataset was prepared by removal of potential false positives as well as including experimentally confirmed inactive molecules rather than seeding the database with decoy molecules (i.e. molecules \textit{assumed} to be inactive).

The targets in the LIT-PCBA datasets have a real-world active-to-inactive molecules ratio, meaning that in some cases the number of confirmed active molecules is strongly imbalanced. 
As ML methods can be severely affected by class imbalance in a dataset \cite{Bishop2006PatternLearning,Murphy2012MachinePerspective}, we balanced active-to-inactive ratios of each target dataset.
However, for some targets in the LIT-PCBA dataset, the number of active molecules is astoundingly small (less than 30), with thousands of confirmed inactive molecules.
In such cases, we padded the dataset with a 1:6 active-to-inactive ratio to balance the dataset (see Appendix~\ref{secA1}). 

Following the above procedure and considerations, we also choose to assess the performance of our method by screening a novel dataset containing known active and inactive COVID-19 inhibitors \citep{Gawriljuk2021MachineSARS-CoV-2}. 
This dataset is particularly challenging for ML classification tasks due to the fact that more than 30\% of the active or inactive molecules do not share a common scaffold or other recurrent structural features, showing how different and diverse the active molecules are, making challenging the classification based on a purely structural basis (i.e.~using a fingerprint-based approach). 

\subsection{Support Vector Classifier and Quantum Kernel}\label{SVC description}

Support Vector Machines (SVMs)~\cite{Jorissen2005VirtualMachine,Sato2012ApplicationInhibitors,Plewczynski2009PerformanceScreening,Lavecchia2015Machine-learningApplications} are among the most widely used supervised machine learning methods for LB-VS. In its basic version, a Support Vector Classifier (SVC)~\cite{cortes_support-vector_1995}, is a SVM-based linear model that, given a training set of the form $\{(\vec{x}_i,y_i) \vert \vec{x}_i \in \mathbb{R}^m, y_i = \pm 1 \}$, finds an optimal separating hyperplane in the data space to distinguish between two given classes, labelled by $y_i$.
Support Vector Machines can be turned into non-linear classifiers by lifting the input data into a higher dimensional feature space $\mathcal{F}$ through a feature map $\phi(\vec{x})$. 
Indeed, while in its elementary form a SVC relies on the computation of similarities between data points through inner products $\langle\vec{x}_i,\vec{x}_j\rangle$ in $\mathbb{R}^m$, in the more general case this is replaced by the corresponding feature space quantity $\langle\phi(\vec{x}_i),\phi(\vec{x}_j)\rangle$, thus allowing for complex non-linear representations of the original inputs. 
The SVC is usually trained by solving a constrained quadratic optimization problem, with a target function of the form
\begin{equation}
    \mathcal{C}(\vec{\alpha}) = \sum_i  \alpha_i - \sum_{ij}y_iy_j\alpha_i\alpha_j\langle\phi(\vec{x}_i),\phi(\vec{x}_j)\rangle
\end{equation}
subject to $\sum_i \alpha_iy_i = 0$ and $0\leq \alpha_i \leq 1/(2nC)$. 
Here $n$ is the cardinality of the training set and $C$ is a regularization parameter controlling the hardness of the margin.
Moreover, the so-called kernel trick~\cite{Aizerman67theoretical} allows one to replace all scalar products with a symmetric positive definite kernel function $k(\vec{x}_i,\vec{x}_j)$, such that, in general, the feature map $\phi$ is only specified in an implicit manner.
Some popular choices for classical SVCs are represented by polynomial kernels $k(\vec{x}_i,\vec{x}_j)=(\vec{x}_i\cdot\vec{x}_j+r)^d$, which reduce to the linear case for $d=1$, and the gaussian radial basis function (RBF) kernel $k(\vec{x}_i,\vec{x}_j)=\exp{(-\gamma\vert\vec{x}_i-\vec{x}_j\vert^2)}$.

The family of quantum kernel methods~\cite{Schuld2019QuantumSpaces,Havlicek2019SupervisedSpaces} employs a quantum computer to extend the class of available kernel functions, particularly to those instances which are believed to be hard to define and compute classically. 
Here, the complex-valued Hilbert space of quantum mechanical states plays the role of the feature space through a mapping of form $\vec{x}\mapsto \vert\phi(\vec{x})\rangle\langle \phi(\vec{x})\vert$ and a natural kernel choice is provided by $k(\vec{x}_i,\vec{x}_j) = \vert\langle\phi(\vec{x}_i)\vert\phi(\vec{x}_j)\rangle\vert^2$. 
The specific nature and properties of the quantum kernel depend on the explicit choice of the embedding function $\vert\phi(\vec{x})\rangle$, which is typically realized as $\vert\phi(\vec{x})\rangle = U(\vec{x})\vert 0\rangle$ with a unitary operation $U$ acting on a reference state $\vert 0\rangle$ and depending parametrically on the classical input data.

Some important remarks are in order here: on the one hand, it is not difficult to see how carefully designed $U$ can in principle lead to classically intractable kernels. On the other hand, it has also been argued that the mere computational hardness of $U$ does not \textit{a priori} guarantee quantum advantage in speed or accuracy for machine learning tasks. In fact, complexity considerations in this context are influenced by both the availability of training data~\cite{Huang2021PowerLearning} and the inherent structure of the data sets~\cite{glick2021covariant,Liu2021ALearning}. Although a rigorous quantum speed-up in supervised learning with quantum kernel methods has been proven theoretically in at least one specific example~\cite{Liu2021ALearning}, the quest for quantum advantage on naturally occurring data sets remains open.
Here we provide strong empirical evidence towards such a goal, working from the assumption that good candidate models for practical quantum advantage are those that, using a classically hard quantum kernel, exhibit superior performances with respect to all classical counterparts in at least some instances of the problem under study.

In the following, we will make use of a unitary template, known as the ZZ feature map with linear entanglement~\cite{Havlicek2019SupervisedSpaces,Qiskit}, to embed classical feature vectors into quantum states. For a 4-dimensional classical input $\vec{x}=(x_0,x_1,x_2,x_3)$ this reads
\begin{equation}
\includegraphics[width=1\columnwidth]{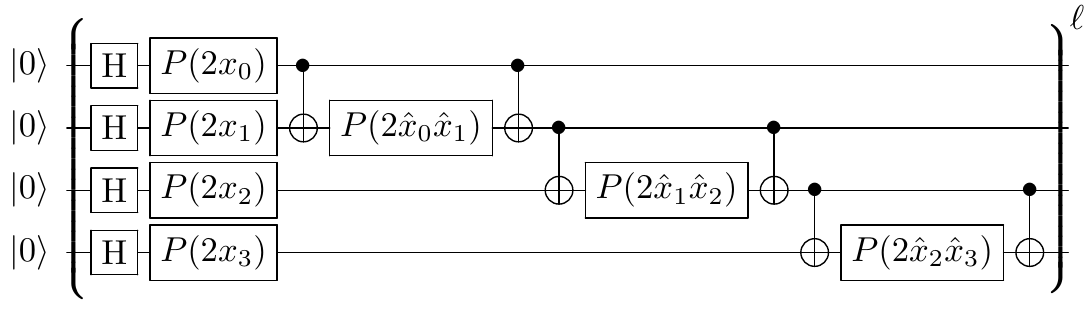}
\label{eq:zz_circuit}
\end{equation}
where $P(\theta)$ is the single-qubit phase gate, $\hat{x}_i\equiv\pi-x_i$ and $\ell$ is called the depth of the feature map. Notice that the number of qubits in the register matches the size of the classical inputs, and the generalization of the parametrized unitary transformation to larger classical inputs is straightforward.
This design is derived from the one presented in Ref.~\cite{Havlicek2019SupervisedSpaces} and yields computationally hard kernels under suitable complexity theory assumptions. In the following, we will set $\ell=2$.

For every pair of classical inputs $\vec{x}$, $\vec{x}'$, the quantum kernel is evaluated on a quantum processor according to the definition
\begin{equation}
    k(\vec{x},\vec{x}') = \vert\langle\phi(\vec{x})\vert\phi(\vec{x}')\rangle\vert^2 = \vert\langle0\vert U^\dagger(\vec{x})U(\vec{x}')\vert 0\rangle\vert^2
\end{equation}
In practice, after initializing the quantum register in the reference state $\vert 0\rangle$, one applies the unitary $U(\vec{x}')$ and the inverse $U^\dagger(\vec{x})=U^{-1}(\vec{x})$, which is easily obtained by reversing the order of operations appearing in Eq.~\eqref{eq:zz_circuit} while also replacing $P(\theta) \rightarrow P(-\theta)$ in all single-qubit phase gates. Finally, the value of the quantum kernel corresponds to the probability of observing the register in the reference state $\vert 0\rangle$ after a measure in the computational basis.

\subsection{Molecular Descriptors} \label{Molecular Descriptors}

Rather than performing VS based on a similarity search of molecules in a database -- using for example molecular fingerprints, like most conventional VS methods -- we trained the classifier using generic features (chemical descriptors), which can be computed for every dataset, regardless of the target type and experimental features.
Chemical descriptors exist in their thousands and can be quickly obtained using specific third-party cheminformatics software tools, both open-source and commercial. 
Our methodology can be coupled with any third-party package capable of producing such descriptors.
Here, we use RDKit~\cite{RDKit:Software} to extract molecular properties starting from molecular structures in a SMILES format. 
As also described in Refs.~\cite{Lo2018MachineDiscovery,Todeschini2000HandbookDescriptors,Bajorath2002IntegrationScreening,Hong2008Mold2Toxicoinformatics,Sawada2014BenchmarkingApproach}, the chosen features can be grouped as 
(\textit{i})~atomic species;
(\textit{ii})~structural properties;
(\textit{iii})~physical-chemical properties;
(\textit{iv})~basic electronic information;
(\textit{v})~molecular complexity – \textit{i.e.} graph descriptors. 
The full list of descriptors can be found in Appendix~\ref{appB}.

\subsection{Descriptors Selection Methods and Quantum Encoding}\label{FeatureSelection}

As described in Sec.~\ref{SVC description}, employing a ZZ feature map implies the use of a parametric type of encoding. Under this scheme, each classical data point must be represented as a vector of real numbers -- in our case the chemical features or molecular descriptors -- that, after a suitable normalization procedure, are used as angles in single-qubit quantum gates (see Eq.~\ref{eq:zz_circuit}).
These operations, in combination with 2-qubit entangling operations, eventually create a quantum superposition state, i.e.~a linear combination of multi-qubit basis states or strings, whose complex coefficients have a non-trivial dependence on the classical input and represent its embedding into the feature Hilbert space.

While in principle very powerful and versatile, this data encoding scheme is rather qubit-intensive, with typically one assigned feature per qubit. As a matter of fact, and despite continued and steady progress in quantum computing technology, implementing a quantum LB-VS workflow featuring hundreds of descriptors is currently demanding even for the largest publicly available quantum processors (127 qubits, IBM Eagle -- late 2021~\cite{ibm_roadmap_2022}). Similarly, this precludes the use of conventional 2048-bit molecular fingerprints, although reduced representations such as the ones proposed by Batra and co-workers~\cite{Batra2021QuantumApplications} may offer viable intermediate-scale solutions.

To achieve a meaningful yet manageable problem size, we therefore extracted, starting from SMILES representations, 47 chemical descriptors as described in Sec.~\ref{Molecular Descriptors}.  
First, each descriptor was normalised using standardization methods.
Second, the descriptors were compressed into $N$ variables using Principal Component Analysis (PCA) \cite{Wold1987PrincipalAnalysis} to match a set of $N$ qubits used by the QSVC algorithm. Alternatively, the best $N$ features were selected using the analysis of variance methodology (ANOVA) \cite{Sthle1989AnalysisANOVA}.
More in detail, PCA -- which is widely adopted for feature reduction in drug discovery \cite{Bajorath2002IntegrationScreening,Elbadawi2021AdvancedDiscovery}~-- was applied via singular value decomposition to obtain a lower-dimensional representation of the data as described in Ref.~\cite{Pedregosa2011Scikit-learn:Python}. Conversely, the ANOVA f-test~\cite{Pedregosa2011Scikit-learn:Python} enabled us to pick the most descriptive $N$ features, as also suggested in Ref.~\cite{Kuhn2019FeatureModels}, leveraging the numerical structure of the inputs and the binary categorical nature (i.e.~active/inactive) of the classification output.
Finally, the resulting feature vectors were used to parametrically initialise the corresponding quantum states on which the quantum kernel entries were evaluated.

\subsection{Evaluation of Performance (ROC)}

The most intuitive way of assessing the performances of binary classification algorithms in both classical and quantum cases is to calculate the AUC-ROC (Receiver Operating Characteristic) metric~\cite{Pedregosa2011Scikit-learn:Python, Fawcett2006AnAnalysis, Hand2001AProblems, McClish1989AnalyzingCurve}. The ROC score provides an indication of the capability of the method to distinguish accurately between true positives and false positives. The score ranges between 0 and 1, where 1 is a perfect classifier capable of retrieving true positives only.   

\section{Results}\label{sec2}

To demonstrate the theoretical benefits of a quantum-inspired LB-VS method, first we ran our experiments simulating noiseless quantum hardware with the python-based Qiskit~\cite{Qiskit} statevector simulator.
Then, we performed simulations introducing statistical and gate noise of actual IBM processors (Qiskit qasm and mock hardware simulators). 
In parallel, we ran identical experiments using the best classical SVC (CSVC) kernel method with optimal hyperparameters (i.e.~the regularization factor $C$ and the RBF kernel non-linearity $\gamma$) obtained via a thorough grid search using the scikit-learn package (see Appendix~\ref{secA1})~\cite{Pedregosa2011Scikit-learn:Python}. To further substantiate our heuristic claims, we performed LB-VS of the same benchmarking datasets using two other state-of-the-art classical methodologies whose performances for generalised VS tasks have been acknowledged in the literature~\cite{Heid2021,Stokes2020,Bartok2017,Nguyen2017,David2020,Coley2020}.
Specifically, we used a Random Matrix Discriminant (RMD) algorithm \cite{Lee2016}, as implemented in PyRMD \cite{Amendola2021PyRMD:Tool}, and a Deep Learning (DL) approach called Directed Message Passing Neural Network (MPNN) \cite{Yang2019}.
We compared the results of the following methodologies ``out-of-the-box'' to have reproducible results:
\begin{itemize}
    \item D-MPNN with 2D CDF-normalized 200 additional features from RDKIT or with binary 2048-bit Morgan fingerprints \cite{Morgan1965}; 
    \item PyRMD with 2048-bit MHFP6 \cite{Probst2018ASettings}, ECFP6 \cite{Rogers2010Extended-ConnectivityFingerprints} and RDKIT fingerprints.
\end{itemize}
Finally, we measured the performance of our QSVC method using actual quantum hardware (IBM Quantum Montreal and IBM Quantum Guadalupe) screening the ADBR2 and the COVID-19 datasets.

\subsection{Results from Numerical Simulations}\label{Results from numerical simulations}

Following the methodology described in Section~\ref{methodology}, all targets of the LIT-PCBA and the COVID-19 datasets were screened using our algorithm. 
To assess the average method performance, each simulation involving both classical and quantum SVC algorithms was run ten times for each feature selection method and per number of features.
Similarly, the RMD and DL classical simulations were also run 10 times.
In the interest of fairness, we quantified the standard deviation of the results using 10-fold cross-validation for all simulations.
In addition, we used identical test and train subsets for all SVC methods, in order to have a true like-for-like comparison between quantum and classical algorithms. 
The full set of numerical results is reported in Tables~\ref{table:ANOVA results} and~\ref{table:PCA results} of Appendix~\ref{LIT appendix}. 

\begin{figure*}[t]
\centering
\includegraphics[width=0.9\textwidth]{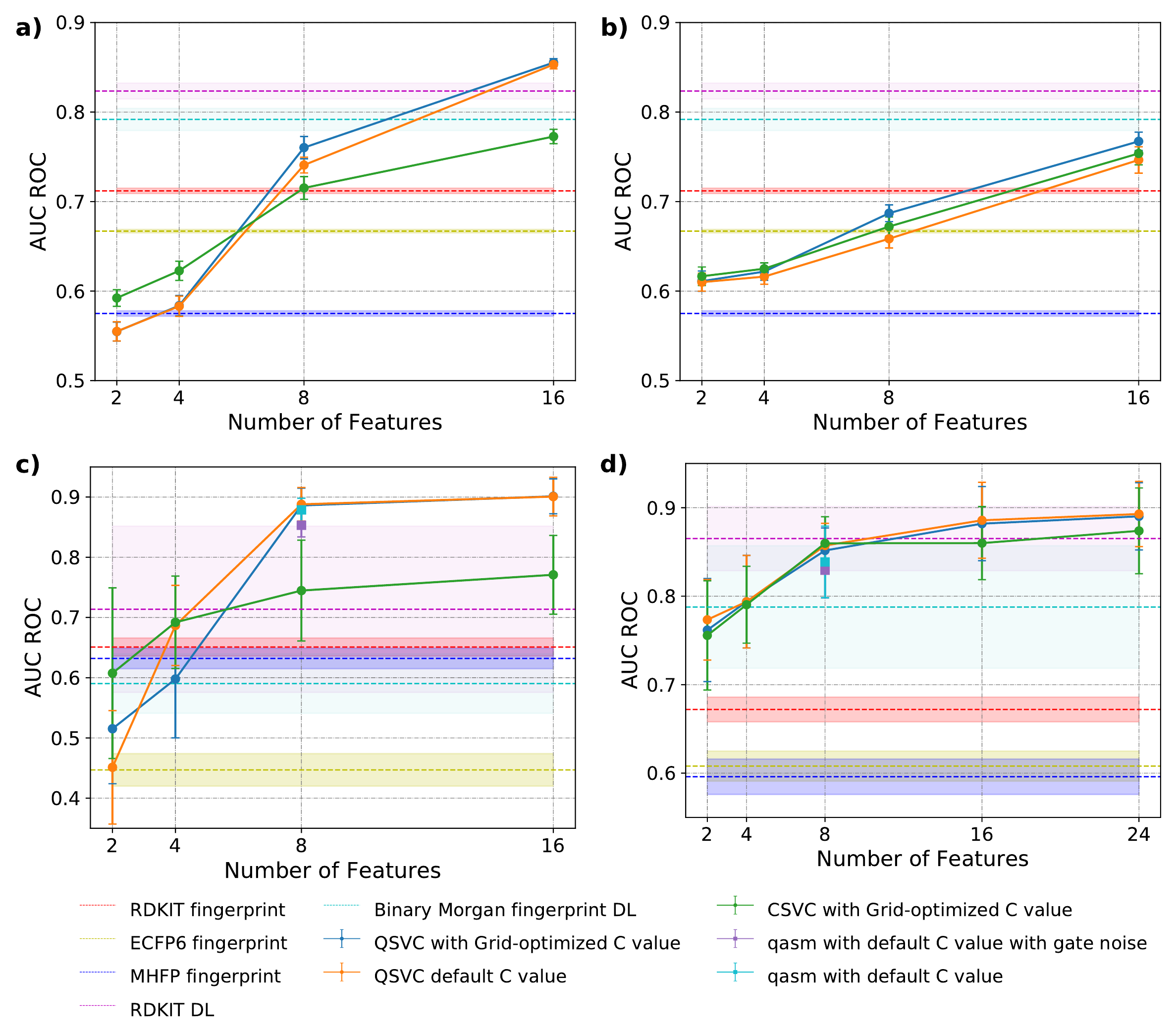}
\centering
\caption{Mean AUC ROC values 
plotted against the number of features used to train the model. a) ALDH1 target with PCA feature reduction method, b) ALDH1 target with ANOVA feature reduction method, c) COVID-19 target with PCA feature reduction method, d) GBA target with ANOVA feature reduction method. In all panels, we compare the results obtained with CSVC, QSVC and the other classical methods, including standard deviations (bars for our CSVC and QSVC, shaded lines for RMD and MPNN). Flat lines represent the mean AUC ROC values of classical methods RMD and MPNN, which are independent of the features number and the features selection method.  %
Panels c) and d) also report average AUC ROC using QSVC algorithm trained with 8 features and using a quantum backend numerical simulator, accounting for statistical noise (Qiskit qasm simulator, cyan) and hardware noise (Montreal backend simulator, purple).}\label{ALDH1}
\end{figure*}

Consistently with classical SVC methods, QSVC statevector results suggest that the overall performance of our method is influenced by \textit{i}) the class balance and the number of the actives in the dataset, \textit{ii}) the feature selection method, \textit{iii})  number of features and iv) the value of regularization parameter $C$.
This trend is captured in Figs.~\ref{histoAppendixANOVA}-\ref{histoAppendixPCA} of Appendix~\ref{LIT appendix}. 
The class balance of the target is the first major factor to affect the performance of our method, similarly to other classical ML/DL approaches.
A small number of active molecules in a dataset (e.g. 17 actives \textit{vs} 312'483 inactives for ADBR2) leads to large standard deviation values (up to ca. $\pm$0.3 of AUC ROC). 
On the other hand, a near negligible standard deviation for target ALDH1, which has the largest number of actives (5'386 actives plus 103'474 inactives), confirms the importance of class balance in terms of stability and performance of all methods. %
The feature selection method is the second major factor to have an effect on the overall performance of all SVC methods described in this paper. 
Due to its very essence, feature selection involves the inevitable loss of information, and this may have a positive or negative effect depending on the dataset/target and the number of features selected. 
Throughout this work, we observe that our QSVC method tends to outperform CSVC and the other classical methods when then number of features is 8 or higher, which also corresponds to a higher number of qubits.
In some cases, such as the ALDH1 target shown in Fig.~\ref{ALDH1}a, there is a near-linear correlation between performance and the number of features/qubits used. 
Finally, the performance of QSVC method is also affected by the choice of the regularization factor $C$. 
This is expected as SVC based methods have an inherent dependency on the $C$ value to determine the misclassification rate of the method. 
Our results suggest that a better QSVC performance is achieved when using a default value~$C=1$.

A more detailed overview is represented in Fig.~\ref{ALDH1}, where the comparison of ANOVA and PCA QSVC(statevector)/CSVC, RMD and DL trends is reported for the ALDH1, COVID-19 (see Sec.~\ref{Dataset}) and GBA (166 active \textit{vs} 296052 inactive molecules) targets. 
Notice that the trend lines of RMD and DL methods are flat since the methods do not make use of an explicit number of features or feature selection methods.
Overall, the performances of both classical and quantum SVC classifiers tend to increase with an incremented number of features used for training, with the steepness of increase and maximum ROC values depending on the features selection method used. 
In other words, the amount of information used to train the algorithm has a major impact on determining the quality of the classifier. 
Fig.~\ref{ALDH1}a shows that the QSVC tends to perform significantly better with 8+ features if compared to its classical equivalent (up to 13\% better when using 16 features), suggesting that the QK method finds a better hyperplane separating active and inactive molecules in the dataset.
As previously discussed, the stability of the method on datasets with optimal class balance such as ALDH1 is suggested by little standard deviation values.
Conversely, class-imbalanced datasets such as COVID-19 and GBA (Fig.~\ref{ALDH1}c-d) typically show higher standard deviations. Nevertheless, the QSVC method consistently keeps higher average accuracy with increasing number of features for at least one choice of the feature selection method (PCA for COVID-19 and ANOVA for GBA).

In Figure~\ref{ALDH1} we also report the mean ROC values obtained following VS of the ALDH1, COVID-19 and GBA targets using the RMD and MPNN methodologies, trained using various fingerprint encoding.
The performance of our quantum classifier seems to outperform the classical VS methodology of RMD with up to ca.~20\%  better mean quantum ROC values when compared to the best value obtained with PyRMD (trained with RDKIT or ECFP6 fingerprints).
Conversely, the MPNN method generally provides better mean ROC values if compared to RMD, and it seems to be outperformed by our methodology only when using a larger number of features.
The standard deviation values of both RMD and MPNN methodologies are in agreement with our CSVC and QSVC methods, confirming the role of class balance in defining the stability of the method. 
In Fig. \ref{ALDH1}c-d we include data points obtained for 8 features using two different quantum simulator backends as implemented in Qiskit, namely the noiseless qasm and the mock Montreal backend simulators.
The former adds statistical noise to the simulation, mimicking the probabilistic quantum measurement process, while the latter simulates both the statistical and hardware noise that one would get on the actual IBM Quantum Montreal processor, whose calibration data are used as parameters in the noise model. 
Due to memory and time limitations, we run simulations with qasm and/or gate noise for 8 features/qubits.
These results show that the addition of noise is not significantly detrimental to the performance of the QSVC method for the COVID-19 and GBA targets.
Furthermore, the standard deviation for the simulations are also in line with the noiseless statevector case. 

Summarizing, the simulation of our quantum LB-VS methodology provides, in a like-for-like comparison, a performance that either compares to other existing classical methodologies or that significantly outperforms them for a sufficiently large number of selected features, sometimes showing evidence of potential linear scale-up. 
This implies that an increase in the number of features/qubits used to train the algorithm can, in principle, further improve the performance of the method. 
Finally, it is important to mention that the prospective quantum advantage suggested by this study is also supported by the geometric test recently introduced in Ref.~\cite{Huang2021PowerLearning}. 
In fact, the data sets shown in Fig.~\ref{ALDH1} noticeably passed such test, yielding $g_{CQ} \propto \sqrt{n}$,  where $g_{CQ}$ is the geometric difference between the quantum kernel and the best classical CSVC counterpart, and $n$ is the number of the data points.

\begin{figure}[b]
\centering
\includegraphics[width=\columnwidth]{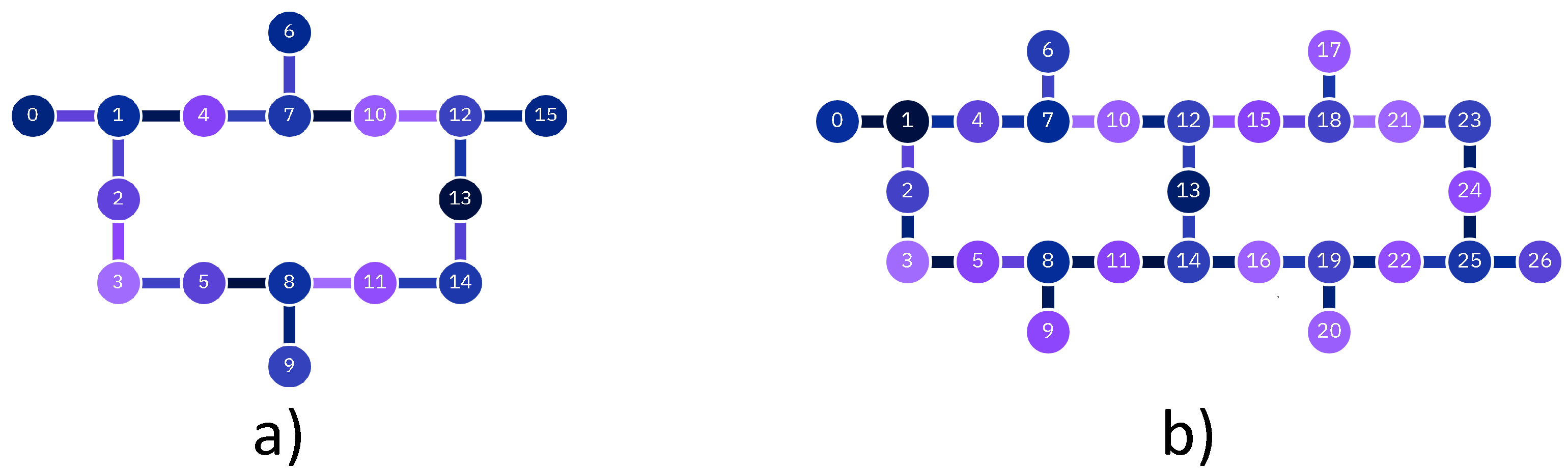}
\caption{IBM Quantum processors layout, a) IBM Quantum Guadalupe : 16 Qubits, 32 QV, 2.4K CLOPS with 
Falcon r4P processor, b) IBM Quantum Montreal: 27 Qubits, 128 QV, 2K CLOPS with 
Falcon r4 processor.}\label{qubits}
\centering
\end{figure}

\begin{figure*}[ht!]
\label{Adbr2}
\centering
\includegraphics[width=\textwidth]{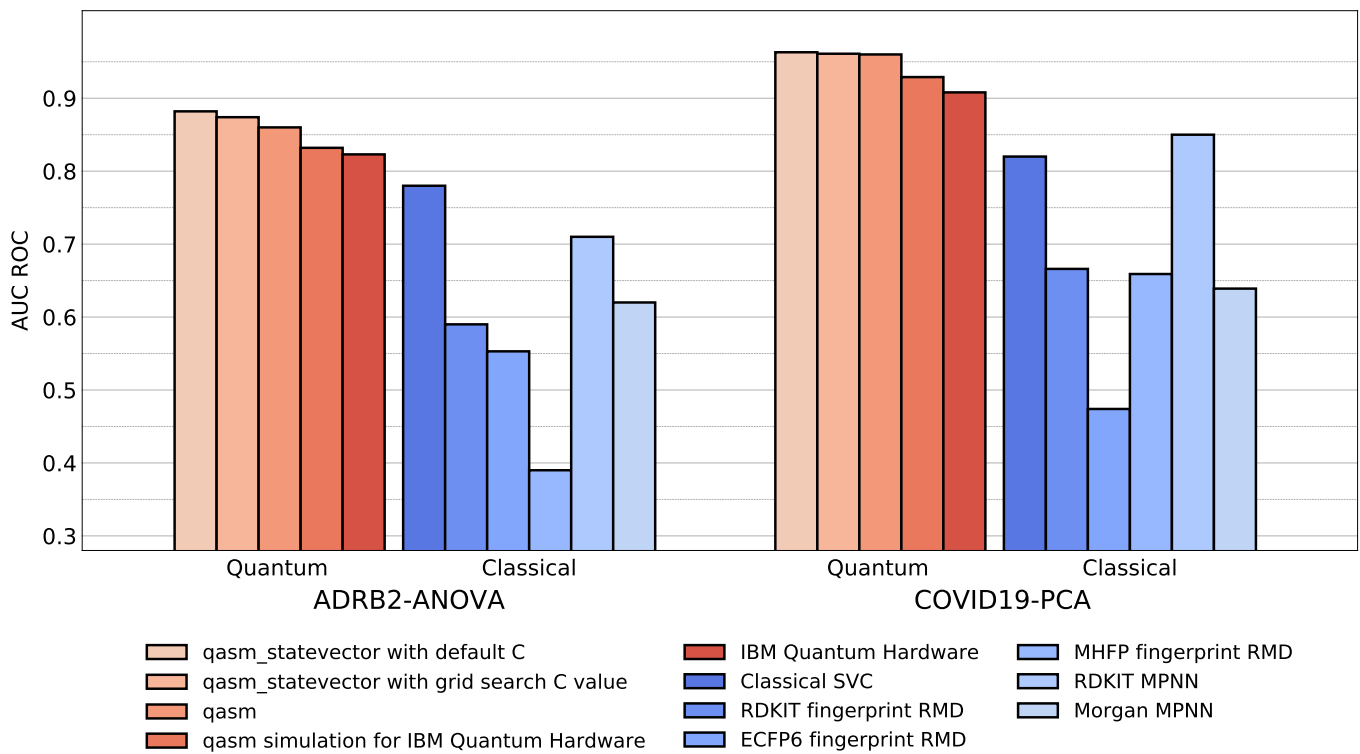}
\caption{Quantum hardware results (single AUC-ROC measurement, in brown), obtained using the IBM Quantum Montreal (ADRB2 - ANOVA, left) and Guadalupe (COVID-19 - PCA, right) quantum processors using 8 qubit and corresponding 8-dimensional feature vectors. Other mean AUC-ROC results were obtained \textit{via} classical and quantum QSVC, RMD and MPNN numerical simulations are also reported, following the trends discussed in Section \ref{Results from numerical simulations}.}\label{hw_results}
\centering
\end{figure*}

\subsection{Results from Quantum Hardware}

In this work, we have so far established that the QSVC method in some cases outperforms an equivalent binary classifier or other classical methods used in VS of databases of molecules, highlighting its dependencies from dataset size, number of features and so forth.
We also showed that the addition of statistical and simulated noise to the quantum algorithm does not considerably reduce the performance of the method, and it lies within the standard deviation of statevector (\textit{i.e.} noise-free) simulations.
The next logical step is therefore to verify whether the quantum algorithm can indeed perform well on actual quantum devices and if the numerical performance is a reliable tool for assessing the potential benefits coming from our method applied to LB-VS on quantum computers.
To this aim, we run simulations on IBM Quantum processors using 8 qubits for ADRB2 on IBM Quantum Montreal (Fig.\ref{qubits}a) and for the COVID-19 dataset on IBM Quantum Guadalupe (Fig.\ref{qubits}b). 

The two datasets underwent the same preparation and simulation methodology as the other datasets described in the previous section.
In Fig.~\ref{hw_results} we report the best results obtained with the optimal feature selection methods (ANOVA for ADRB2, PCA for COVID-19) along with qasm, mock hardware (Montreal and Guadalupe backend) and hardware results.
Also in this case, QSVC statevector numerical simulations for both ADRB2 and COVID-19 datasets provided better results if compared to classical equivalents, including RMD and MPNN methods.
Remarkably, in our experiments we found that training and testing our QSVC algorithm on IBM Quantum Montreal provided an AUC ROC result for ADRB2 comparable to the results of all the other (quantum) numerical simulation methods, with outcomes lying well within the error bar of statevector simulations (see Table~\ref{table:ANOVA results}). 
Similarly, running the QSVC algorithm on IBM Quantum Guadalupe for the COVID-19 dataset also provided an AUC ROC that is comparable to the best (quantum) numerical simulations.
Most importantly, quantum hardware results for both targets greatly outperform the other classical methodologies, thus confirming that quantum kernel methods are indeed a well suited tool for potential active/inactive molecules classification against ADRB2 and COVID-19 targets.

We can conclude that, for these two instances, simulated backends have correctly predicted the respective hardware results, hence providing a truthful insight into what to expect from real quantum hardware. 
These results are encouraging, suggesting that although it might not yet be possible to perform large scale simulations on currently available quantum hardware, promising proof-of-principle demonstrations can be achieved, confirming the reliability and predictive power of numerical simulations.

\section{Discussion and Conclusions}\label{sec12}
In this paper, we have shown that, when using the same data sets and training conditions, a quantum classifier can in some cases outperform the best available classical counterparts.
This implies that the analysis of cheminformatics data can already benefit from the application of quantum technologies, which have the potential of enhancing the success rate of, e.g., structure classification for material design and drug discovery. 
We need to stress that for this particular investigation, we reduced the problem size (number of features) to a level that is currently affordable for quantum calculations, using both classical simulations of quantum algorithms and the execution on state-of-the-art quantum hardware. 
This implies for the moment the use of relatively small data sets and the selection of a limited number of representative features for the characterization of the compound properties. 
In doing that, the proposed quantum classifiers are still classically accessible in simulations, hence our approach cannot yet lead to direct quantum advantage -- at best, it can be qualified as an interesting new quantum-inspired classical algorithm.
However, and this is the main message of this paper, there are currently no evident reasons to doubt that the same advantage observed in the proof-of-principle applications exposed in this work will not withstand the scaling-up to a larger number of descriptors (features). 
These will correspond to a number of qubits (more than 30-50) that will remain inaccessible to classical simulators while becoming manageable on the near-term quantum computers of the coming generations~\cite{ibm_roadmap_2022}. 
To refer to the class of algorithms featuring this property, we introduce the concept of Prospective Quantum Advantage (PQA), as an impactful intermediate step -- with potential value for small and intermediate size instances -- towards full quantum advantage in large scale applications. 
In this sense, we conclude that quantum information algorithms have already an attractive potential for the chemical and pharmaceutical industry, as they bring new value to research workflows and provide solutions superior to previously accessible classical~ones.

\begin{acknowledgments}
This work was supported by the Hartree National Centre for Digital Innovation, a UK Government-funded collaboration between STFC and IBM. IBM, the IBM logo, and ibm.com are trademarks of International Business Machines Corp., registered in many jurisdictions worldwide. Other product and service names might be trademarks of IBM or other companies. The current list of IBM trademarks is available at https://www.ibm.com/legal/copytrade.

\end{acknowledgments}

\onecolumngrid

\bibliography{references}

\clearpage

\appendix

\section{Algorithm Workflow}\label{secA1}

\begin{figure}[t]
\centering
\includegraphics[width=1\textwidth]{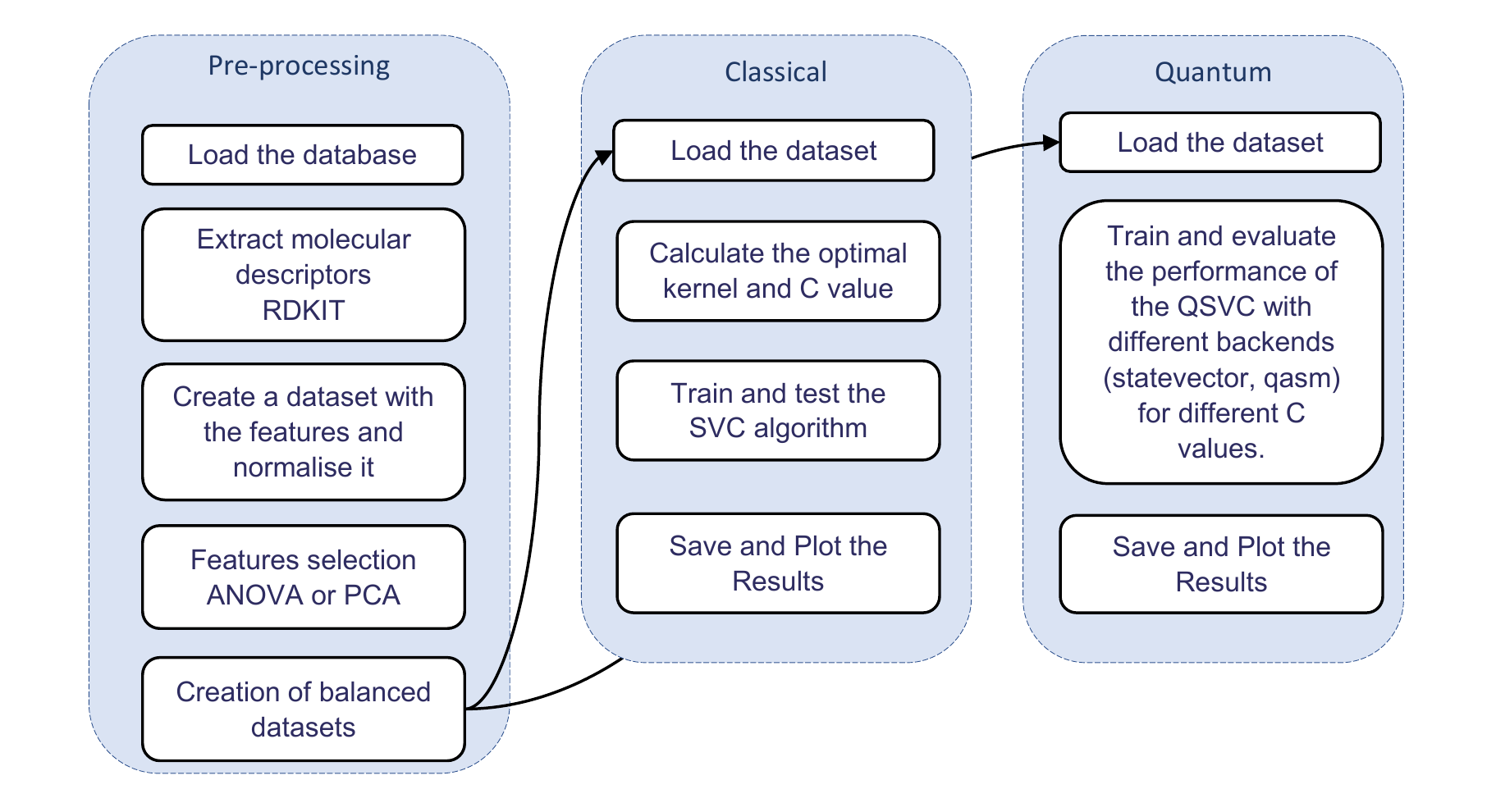}
\caption{Schematic representation of the algorithm workflow. In light blue are represented the three main steps of the workflow: preprocessing, classical SVC and Quantum SVC.}
\label{fig:app_workflow}
\end{figure}

The algorithm is divided as follows (see also Fig.~\ref{fig:app_workflow}).
\begin{enumerate}
    \item{\textit{Preprocessing step.} The dataset of molecules (encoded in a SMILES format) is loaded and 47 molecular descriptors are extracted using RDKit.
    A new dataset (80\% training and 20\% validating)  is then created and normalised, containing the extracted features per molecule. 
    A novel dataset is created for each new set of reduced features and per reduction method. 
    Each dataset is then balanced to have equal amounts of actives and inactive molecules or, when the number of actives is below 30, the number of inactive is padded to six times the number of actives.}
    \item{\textit{Classical SVC step.}  Using the training portion, a Grid-Search is performed over $C$ and $\gamma$ parameters as well as kernel type (Poly, RFB, Linear). 
    The best C, $\gamma$ and kernel type are then used to train the SVC algorithm.
    Therefore, the algorithm is validated over the validating portion and the AUC ROC is measured.} 
    \item{\textit{Quantum SVC.} The training set for each reduced features dataset obtained in step 2 is re-used to compare Quantum and Classical SVC results. Here, the training is performed by the calculation of a Quantum Kernel via the ZZ feature Map. Two computations are run in parallel. The first uses the same $C$ and $\gamma$ parameters obtained in step 2 and used by the Classical SVC algorithm (a like-for-like comparison). The second uses default Qiskit $C$ and $\gamma$ values. Validation of algorithm performance follows the training and AUC ROC is measured.
    Steps 2 and 3 have been repeated ten times per dataset to verify the stability of our method.
    }
\end{enumerate}

\section{Molecular Descriptors}\label{appB}
All the molecular descriptors used in our work are presented in Table~\ref{tab:descriptors}.

\begin{table}[!]
\caption{Table containing all the molecular descriptors used in this work. All the descriptors were extracted using the RDKit package.}
\label{tab:descriptors}
\vspace{3.0ex}
\begin{adjustbox}{max width=\textwidth, center}
\maxsizebox{7cm}{!}{
\begin{tabular}{l|r}
\hline \\ [0.5ex] 
\textbf{Category}                                        & \textbf{Molecular descriptor}                     \\
                                                         &                                                   \\
                                                         \hline                                              \\
\multirow{9}{*}{\textbf{Atomic species}}                 & C                                                 \\
                                                         & N                                                 \\
                                                         & O                                                 \\
                                                         & P                                                 \\
                                                         & S                                                 \\
                                                         & F                                                 \\
                                                         & Cl                                                \\
                                                         & Br                                                \\
                                                         & I                                                 \\
                                                         &                                                   \\
                                                         \hline                                              \\
\multirow{11}{*}{\textbf{Structural   properties}}       & Single\_Bonds                                     \\
                                                         & Double\_Bonds                                     \\
                                                         & NumStereoE                                        \\
                                                         & Num\_Aromatic\_Atoms                              \\
                                                         & Aromatic\_Proportion                              \\
                                                         & NumRotatableBonds                                 \\
                                                         & Total\_NH\_OH                                     \\
                                                         & Total\_N\_O                                       \\
                                                         & NumHydrogenAcceptors                              \\
                                                         & NumHydrogenDonors                                 \\
                                                         & NumofHeteroatoms                                  \\
                                                         &                                                   \\
                                                         \hline                                              \\
\multirow{2}{*}{\textbf{Physical-chemical properties}}   & MolLogP                                           \\
                                                         & MolWt                                             \\
                                                         &                                                   \\
                                                         \hline                                              \\
\multirow{6}{*}{\textbf{Basic electronic   information}} & FpDensityMorgan1                                  \\
                                                         & FpDensityMorgan2                                  \\
                                                         & FpDensityMorgan3                                  \\
                                                         & MaxAbsPartialCharge                               \\
                                                         & MinAbsPartial Charge                           \\
                                                         & NumValenceElectrons                               \\
                                                         &                                                   \\
                                                         \hline                                              \\
\multirow{19}{*}{\textbf{Molecular complexity}}          & BertzCT                                           \\
                                                         & BalabanJ                                          \\
                                                         & Chi0                                              \\
                                                         & Chi0n                                             \\
                                                         & Chi0v                                             \\
                                                         & Chi1                                              \\
                                                         & Chi1n                                             \\
                                                         & Chi1v                                             \\
                                                         & Chi2n                                             \\
                                                         & Chi2v                                             \\
                                                         & Chi3n                                             \\
                                                         & Chi3v                                             \\
                                                         & Chi4n                                             \\
                                                         & Chi4v                                             \\
                                                         & HallKierAlpha                                     \\
                                                         & Ipc                                               \\
                                                         & Kappa1                                            \\
                                                         & Kappa2                                            \\
                                                         & Kappa3                                            \\
                                                         \hline \\
\end{tabular}
}
\end{adjustbox}
\end{table}

\section{LIT-PCBA Screening Results}\label{LIT appendix}

In this section, we collect numerical results for all LIT-PBA sub-sets and COVID-19 dataset, calculated using CSVC, QSVC via Qiskit statevector simulator, PyRMD and MPNN. Results are divided by features selection method (ANOVA/PCA). 

\begin{sidewaystable}[!]
\caption{Average AUC ROC values for 15 targets of the LIT-PCBA dataset and the COVID-19 dataset, obtained using our classical and quantum SVC method (with \textbf{ANOVA} features selection) and fingerprint similarity using PyRMD.}
\begin{adjustbox}{width=\textwidth}
  \centering
    \begin{tabular}{lrrrrr|rrrrr|rrrrr|rrr|rr}
    \hline \\ [0.5ex] 
& \multicolumn{5}{@{}c@{}}{\textbf{CSVC}}& \multicolumn{5}{@{}c@{}}{\textbf{QSVC ANOVA (Default C)}}&\multicolumn{5}{@{}c@{}}{\textbf{QSVC ANOVA (CSVC C values)}}&\multicolumn{3}{@{}c@{}}{\textbf{PyRMD}}& \multicolumn{2}{@{}c@{}}{\textbf{ChemProp (MPNN)}}\\
\textbf{Target} & 2	& 4 & 8 & 16 & 24 & 2 & 4 & 8 & 16 & 24 & 2	& 4 & 8 & 16 & 24 & RDKIT & ECFP6 & MHFP & RDKIT & Morgan \\
\midrule
        \textbf{ADRB2} & 0.508($\pm$ 0.223) & 0.661($\pm$ 0.172) & 0.652($\pm$ (0.138) & 0.632($\pm$ 0.228) & 0.549($\pm$ 0.214) & 0.465($\pm$ 0.305) & 0.732($\pm$ 0.127) & 0.777($\pm$ 0.0874) & 0.815($\pm$ 0.0944) & 0.546($\pm$ 0.302) & 0.507($\pm$ 0.242) & 0.759($\pm$ 0.109) & 0.746($\pm$ 0.121) & 0.812($\pm$ 0.089) & 0.617($\pm$ 0.282) & 0.524($\pm$ 0.061) & 0.453($\pm$ 0.057) & 0.359($\pm$ 0.031)&0.498($\pm$ 0.218)&0.497($\pm$ 0.127) \\
        \textbf{ALDH1} & 0.617($\pm$ 0.01) & 0.625($\pm$ 0.007) & 0.672($\pm$ 0.011) & 0.754($\pm$ 0.013) & n/a & 0.61($\pm$ 0.01) & 0.616($\pm$ 0.009) & 0.659($\pm$ 0.01) & 0.747($\pm$ 0.015) & n/a & 0.611($\pm$ 0.011) & 0.622($\pm$ 0.01) & 0.687($\pm$ 0.009) & 0.767($\pm$ 0.01) & n/a & 0.712($\pm$ 0.003) & 0.667($\pm$ 0.002) & 0.575($\pm$ 0.003)& 0.824($\pm$0.009)& 0.792($\pm$0.012)  \\ 
        \textbf{ESR1}\textbf{\_ago} & 0.346($\pm$ 0.210) & 0.425($\pm$ 0.278) & 0.557($\pm$ 0.227) & 0.497($\pm$ 0.234) & 0.486($\pm$ 0.177) & 0.362($\pm$ 0.267) & 0.382($\pm$ 0.155) & 0.415($\pm$ 0.145) & 0.402($\pm$ 0.18) & 0.493($\pm$ 0.22) & 0.406($\pm$ 0.211) & 0.455($\pm$ 0.269) & 0.488($\pm$ 0.174) & 0.414($\pm$ 0.164) & 0.483($\pm$ 0.231) & 0.528($\pm$ 0.041) & 0.496($\pm$ 0.05) & 0.585($\pm$ 0.026)& 0.618($\pm$0.315) & 0.344($\pm$0.203)  \\ 
        \textbf{ESR1}\textbf{\_ant} & 0.646($\pm$ 0.109) & 0.636($\pm$ 0.072) & 0.715($\pm$ 0.046) & 0.688($\pm$ 0.037) & 0.729($\pm$ 0.055) & 0.693($\pm$ 0.086) & 0.635($\pm$ 0.076) & 0.78($\pm$ 0.041) & 0.816($\pm$ 0.033) & 0.822($\pm$ 0.047) & 0.689($\pm$ 0.088) & 0.557($\pm$ 0.108) & 0.75($\pm$ 0.059) & 0.797($\pm$ 0.047) & 0.818($\pm$ 0.048) & 0.661($\pm$ 0.013) & 0.477($\pm$ 0.022) & 0.596($\pm$ 0.011)&  0.768($\pm$0.071) & 0.700($\pm$0.104)\\ 
        \textbf{FEN1} & 0.698($\pm$ 0.038) & 0.787($\pm$ 0.027) & 0.835($\pm$ 0.024) & 0.843($\pm$ 0.028) & 0.880($\pm$ 0.031) & 0.705($\pm$ 0.044) & 0.795($\pm$ 0.031) & 0.826($\pm$ 0.027) & 0.855($\pm$ 0.016) & 0.871($\pm$ 0.03) & 0.697($\pm$ 0.038) & 0.772($\pm$ 0.034) & 0.761($\pm$ 0.047) & 0.85($\pm$ 0.016) & 0.869($\pm$ 0.028) & 0.755($\pm$ 0.012) & 0.735($\pm$ 0.01) & 0.686($\pm$ 0.01)&0.952($\pm$0.014) & 0.893($\pm$0.015)  \\ 
        \textbf{GBA} & 0.756($\pm$ 0.062) & 0.790($\pm$ 0.043) & 0.859($\pm$ 0.030) & 0.860($\pm$ 0.041) & 0.874($\pm$ 0.048) & 0.773($\pm$ 0.0455) & 0.794($\pm$ 0.052) & 0.857($\pm$ 0.025) & 0.886($\pm$ 0.043) & 0.893($\pm$ 0.037) & 0.762($\pm$ 0.058) & 0.794($\pm$ 0.052) & 0.852($\pm$ 0.025) & 0.882($\pm$ 0.042) & 0.890($\pm$ 0.038) & 0.672($\pm$ 0.014) & 0.608($\pm$ 0.017) & 0.596($\pm$ 0.02) & 0.865($\pm$0.036) & 0.788($\pm$0.069) \\ 
        \textbf{IDH1} & 0.491($\pm$ 0.138) & 0.482($\pm$ 0.134) & 0.553($\pm$ 0.084) & 0.752($\pm$ 0.085) & 0.708($\pm$ 0.081) & 0.623($\pm$ 0.133) & 0.563($\pm$ 0.168) & 0.658($\pm$ 0.088) & 0.427($\pm$ 0.16) & 0.678($\pm$ 0.08) & 0.628($\pm$ 0.166) & 0.547($\pm$ 0.17) & 0.557($\pm$ 0.161) & 0.426($\pm$ 0.141) & 0.673($\pm$ 0.086) & 0.692($\pm$ 0.015) & 0.484($\pm$ 0.028) & 0.504($\pm$ 0.031)& 0.725($\pm$0.12) & 0.669($\pm$0.113)  \\ 
        \textbf{KAT2A} & 0.587($\pm$ 0.031) & 0.629($\pm$ 0.043) & 0.647($\pm$ 0.053) & 0.684($\pm$ 0.064) & 0.653($\pm$ 0.056) & 0.594($\pm$ 0.025) & 0.621($\pm$ 0.051) & 0.666($\pm$ 0.062) & 0.683($\pm$ 0.047) & 0.616($\pm$ 0.044) & 0.588($\pm$ 0.028) & 0.57($\pm$ 0.094) & 0.62($\pm$ 0.054) & 0.684($\pm$ 0.034) & 0.624($\pm$ 0.039) & 0.568($\pm$ 0.022) & 0.546($\pm$ 0.014) & 0.475($\pm$ 0.011)& 0.729($\pm$0.048)& 0.636($\pm$0.098)  \\ 
        \textbf{MAPK1} & 0.693($\pm$ 0.052) & 0.71($\pm$ 0.028) & 0.72($\pm$ 0.033) & 0.716($\pm$ 0.026) & 0.699($\pm$ 0.039) & 0.674($\pm$ 0.064) & 0.690($\pm$ 0.034) & 0.741($\pm$ 0.043) & 0.738($\pm$ 0.034) & 0.667($\pm$ 0.024) & 0.675($\pm$ 0.059) & 0.653($\pm$ 0.056) & 0.717($\pm$ 0.071) & 0.73($\pm$ 0.042) & 0.663($\pm$ 0.025) & 0.578($\pm$ 0.006) & 0.565($\pm$ 0.01) & 0.524($\pm$ 0.009)& 0.768($\pm$0.043) & 0.709($\pm$0.057)  \\ 
        \textbf{MTORC1} & 0.586($\pm$ 0.077) & 0.611($\pm$ 0.098) & 0.702($\pm$ 0.079) & 0.652($\pm$ 0.068) & 0.678($\pm$ 0.077) & 0.575($\pm$ 0.119) & 0.633($\pm$ 0.045) & 0.628($\pm$ 0.085) & 0.604($\pm$ 0.14) & 0.600($\pm$ 0.112) & 0.613($\pm$ 0.074) & 0.574($\pm$ 0.097) & 0.615($\pm$ 0.081) & 0.605($\pm$ 0.14) & 0.609($\pm$ 0.095) & 0.62($\pm$ 0.014) & 0.576($\pm$ 0.02) & 0.547($\pm$ 0.015) & 0.729($\pm$0.102) & 0.682($\pm$0.13)  \\
        \textbf{OPRK1} & 0.801($\pm$ 0.07) & 0.809($\pm$ 0.116) & 0.897($\pm$ 0.04) & 0.887($\pm$ 0.101) & 0.914($\pm$ 0.087) & 0.693($\pm$ 0.184) & 0.693($\pm$ 0.192) & 0.789($\pm$ 0.125) & 0.867($\pm$ 0.095) & 0.93($\pm$ 0.053) & 0.789($\pm$ 0.094) & 0.771($\pm$ 0.157) & 0.866($\pm$ 0.043) & 0.891($\pm$ 0.079) & 0.929($\pm$ 0.052) & 0.778($\pm$ 0.024) & 0.594($\pm$ 0.046) & 0.55($\pm$ 0.007) & 0.790($\pm$0.246) & 0.728($\pm$0.275) \\
        \textbf{PKM2} & 0.729($\pm$0.033) & 0.738($\pm$0.033) & 0.747($\pm$0.028) & 0.788($\pm$0.035) & 0.785($\pm$0.037) & 0.73($\pm$0.029) & 0.731($\pm$0.032) & 0.737($\pm$0.028) & 0.783($\pm$0.024) & 0.761($\pm$0.022) & 0.594($\pm$0.043) & 0.654($\pm$0.03) & 0.677($\pm$0.034) & 0.742($\pm$0.022) & 0.767($\pm$0.019)& 0.678($\pm$ 0.005) & 0.647($\pm$ 0.008) & 0.55($\pm$ 0.007) & 0.813($\pm$0.036) & 0.786($\pm$0.042)\\ 
        \textbf{PPARG} & 0.508($\pm$0.194) & 0.702($\pm$0.07) & 0.829($\pm$0.052) & 0.767($\pm$0.075) & 0.836($\pm$0.035) & 0.544($\pm$0.265) & 0.701($\pm$0.078) & 0.775($\pm$0.072) & 0.749($\pm$0.06) & 0.666($\pm$0.197) & 0.665($\pm$0.157) & 0.692($\pm$0.056) & 0.794($\pm$0.055) & 0.739($\pm$0.059) & 0.661($\pm$0.195) & 0.654($\pm$ 0.023) & 0.701($\pm$ 0.019) & 0.692($\pm$ 0.018)& 0.837($\pm$0.209)& 0.728($\pm$0.211)\\ 
        \textbf{TP53} & 0.64($\pm$ 0.14) & 0.654($\pm$ 0.101) & 0.725($\pm$ 0.102) & 0.756($\pm$ 0.082) & 0.766($\pm$ 0.063) & 0.693($\pm$ 0.104) & 0.649($\pm$ 0.088) & 0.827($\pm$ 0.078) & 0.858($\pm$ 0.035) & 0.844($\pm$ 0.052) & 0.66($\pm$ 0.118) & 0.541($\pm$ 0.144) & 0.826($\pm$ 0.076) & 0.849($\pm$ 0.044) & 0.849($\pm$ 0.048) & 0.609($\pm$ 0.026) & 0.536($\pm$ 0.02) & 0.503($\pm$ 0.018) & 0.717($\pm$0.144) & 0.712($\pm$0.116)\\ 
        \textbf{VDR} & 0.684($\pm$0.03) & 0.683($\pm$0.032) & 0.709($\pm$0.034) & 0.765($\pm$0.027) & 0.808($\pm$0.03) & 0.675($\pm$0.026) & 0.684($\pm$0.03) & 0.7($\pm$0.036) & 0.75($\pm$0.023) & 0.754($\pm$0.027) & 0.68($\pm$0.027) & 0.676($\pm$0.032) & 0.684($\pm$0.036) & 0.733($\pm$0.028) & 0.755($\pm$0.018) & 0.747($\pm$ 0.009) & 0.674($\pm$ 0.009) & 0.634($\pm$ 0.007) & 0.888($\pm$0.027) & 0.829($\pm$0.028) \\\hline
        \textbf{COVID-19} & 0.476($\pm$0.091) & 0.613($\pm$0.147) & 0.65($\pm$0.068) & 0.693($\pm$0.046) & n/a	& 0.517($\pm$0.137) & 0.639($\pm$0.129) & 0.683($\pm$0.058) & 0.804($\pm$0.067) & n/a & 0.532($\pm$0.096) & 0.703($\pm$0.063) & 0.700($\pm$0.088) & 0.755($\pm$0.09) & n/a & 0.651($\pm$0.015)	& 0.447($\pm$0.027) & 0.632($\pm$0.017) & 0.713($\pm$0.137) & 0.590($\pm$0.049)\\
    \hline \\ [0.5ex]
    \end{tabular}%
\end{adjustbox}
\label{table:ANOVA results}
\end{sidewaystable}

\begin{sidewaystable}[!]
\caption{Average AUC ROC values for 15 targets of the LIT-PCBA dataset and the COVID-19 dataset, obtained using our classical and quantum SVC method (with \textbf{PCA} features selection) and fingerprint similarity using PyRMD and ChemProp (MPNN).}
\begin{adjustbox}{width=\textwidth}
  \centering
    \begin{tabular}{lrrrrr|rrrrr|rrrrr|rrr|rr}
    \hline \\ [0.5ex] 
& \multicolumn{5}{@{}c@{}}{\textbf{CSVC}}& \multicolumn{5}{@{}c@{}}{\textbf{QSVC PCA (Default C)}}&\multicolumn{5}{@{}c@{}}{\textbf{QSVC PCA (CSVC C values)}}&\multicolumn{3}{@{}c@{}}{\textbf{PyRMD}}&\multicolumn{2}{@{}c@{}}{\textbf{ChemProp (MPNN)}}  \\
\textbf{Target} & 2	& 4 & 8 & 16 & 24 & 2 & 4 & 8 & 16 & 24 & 2	& 4 & 8 & 16 & 24 & RDKIT & ECFP6 & MHFP & RDKIT & Morgan\\
\midrule
        \textbf{ADRB2} & 0.483($\pm$0.168) & 0.499($\pm$0.181) & 0.752($\pm$0.188) & 0.634($\pm$0.234) & 0.682($\pm$0.115) & 0.412($\pm$0.122) & 0.435($\pm$0.175) & 0.364($\pm$0.124) & 0.380($\pm$0.235) & 0.424($\pm$0.297) & 0.436($\pm$0.096) & 0.440($\pm$0.198) & 0.370($\pm$0.124) & 0.380($\pm$0.235) & 0.425($\pm$0.297) & 0.524($\pm$0.061) & 0.453($\pm$0.057) & 0.359($\pm$0.031)& 0.498($\pm$0.218) & 0.497($\pm$0.127)  \\ 
        \textbf{ALDH1} & 0.592($\pm$0.009) & 0.623($\pm$0.011) & 0.715($\pm$0.013) & 0.773($\pm$0.008) & n/a & 0.555($\pm$0.011) & 0.583($\pm$0.011) & 0.741($\pm$0.009) & 0.853($\pm$0.005) & n/a & 0.555($\pm$0.011) & 0.584($\pm$0.011) & 0.76($\pm$0.012) & 0.855($\pm$0.004) & n/a & 0.712($\pm$0.003) & 0.667($\pm$0.002) & 0.575($\pm$0.003) & 0.824($\pm$0.009) & 0.792($\pm$0.012) \\ 
        \textbf{ESR1\_ago} & 0.519($\pm$0.207) & 0.458($\pm$0.224) & 0.607($\pm$0.161) & 0.549($\pm$0.168) & 0.44($\pm$0.154) & 0.486($\pm$0.199) & 0.457($\pm$0.164) & 0.591($\pm$0.196) & 0.539($\pm$0.133) & 0.585($\pm$0.200) & 0.470($\pm$0.216) & 0.485($\pm$0.169) & 0.581($\pm$0.204) & 0.526($\pm$0.136) & 0.585($\pm$0.200) & 0.528($\pm$0.041) & 0.496($\pm$0.05) & 0.585($\pm$0.026)& 0.618($\pm$0.315)	& 0.344($\pm$0.203)  \\ 
        \textbf{ESR1\_ant} & 0.603($\pm$0.086) & 0.707($\pm$0.06) & 0.79($\pm$0.073) & 0.819($\pm$0.065) & 0.813($\pm$0.062) & 0.463($\pm$0.069) & 0.618($\pm$0.122) & 0.734($\pm$0.061) & 0.763($\pm$0.054) & 0.824($\pm$0.048) & 0.489($\pm$0.067) & 0.620($\pm$0.122) & 0.735($\pm$0.061) & 0.762($\pm$0.054) & 0.824($\pm$0.048) & 0.661($\pm$0.013) & 0.477($\pm$0.022) & 0.596($\pm$0.011)& 0.768($\pm$0.071) &	0.700($\pm$0.104)  \\ 
        \textbf{FEN1} & 0.619($\pm$0.042) & 0.808($\pm$0.025) & 0.835($\pm$0.029) & 0.888($\pm$0.026) & 0.915($\pm$0.027) & 0.612($\pm$0.025) & 0.629($\pm$0.051) & 0.763($\pm$0.019) & 0.799($\pm$0.024) & 0.809($\pm$0.035) & 0.609($\pm$0.025) & 0.629($\pm$0.051) & 0.763($\pm$0.019) & 0.798($\pm$0.024) & 0.810($\pm$0.035) & 0.755($\pm$0.012) & 0.735($\pm$0.01) & 0.686($\pm$0.01) & 0.952($\pm$0.014) &	0.893($\pm$0.015)  \\ 
        \textbf{GBA} & 0.756($\pm$0.052) & 0.785($\pm$0.044) & 0.882($\pm$0.033) & 0.899($\pm$0.032) & 0.900($\pm$0.028) & 0.717($\pm$0.054) & 0.654($\pm$0.054) & 0.781($\pm$0.047) & 0.804($\pm$0.039 & 0.831($\pm$0.034) & 0.717($\pm$0.054) & 0.660($\pm$0.047) & 0.781($\pm$0.046) & 0.805($\pm$0.037) & 0.831($\pm$0.034) & 0.672($\pm$0.014) & 0.608($\pm$0.017) & 0.596($\pm$0.02) & 0.865($\pm$0.036)	& 0.788($\pm$0.069) \\ 
        \textbf{IDH1} & 0.523($\pm$0.084) & 0.563($\pm$0.074) & 0.718($\pm$0.082) & 0.728($\pm$0.069) & 0.824($\pm$0.078) & 0.505($\pm$0.096) & 0.507($\pm$0.152) & 0.616($\pm$0.127) & 0.705($\pm$0.082) & 0.575($\pm$0.180) & 0.539($\pm$0.074) & 0.509($\pm$0.119) & 0.595($\pm$0.123) & 0.701($\pm$0.08) & 0.573($\pm$0.178) & 0.692($\pm$0.015) & 0.484($\pm$0.028) & 0.504($\pm$0.031) & 0.725($\pm$0.12) & 	0.669($\pm$0.113)\\ 
        \textbf{KAT2A} & 0.532($\pm$0.081) & 0.579($\pm$0.051) & 0.627($\pm$0.095) & 0.677($\pm$0.044) & 0.668($\pm$0.054) & 0.478($\pm$0.054) & 0.464($\pm$0.061) & 0.602($\pm$0.051) & 0.627($\pm$0.043) & 0.632($\pm$0.06) & 0.489($\pm$0.064) & 0.53($\pm$0.092) & 0.601($\pm$0.052) & 0.627($\pm$0.043) & 0.632($\pm$0.061) & 0.568($\pm$0.022) & 0.546($\pm$0.014) & 0.475($\pm$0.011)& 0.729($\pm$0.048) & 0.636($\pm$0.098) \\ 
        \textbf{MAPK1} & 0.675($\pm$0.055) & 0.682($\pm$0.067) & 0.701($\pm$0.032) & 0.728($\pm$0.041) & 0.741($\pm$0.038) & 0.615($\pm$0.042) & 0.630($\pm$0.048) & 0.600($\pm$0.035) & 0.633($\pm$0.038) & 0.654($\pm$0.046) & 0.611($\pm$0.046) & 0.591($\pm$0.069) & 0.599($\pm$0.033) & 0.633($\pm$0.038) & 0.651($\pm$0.045) & 0.578($\pm$0.006) & 0.565($\pm$0.01) & 0.524($\pm$0.009) & 0.768($\pm$0.043) &	0.709($\pm$0.057) \\ 
        \textbf{MTORC1} & 0.559($\pm$0.12) & 0.557($\pm$0.077) & 0.641($\pm$0.072) & 0.667($\pm$0.075) & 0.677($\pm$0.076) & 0.437($\pm$0.055) & 0.418($\pm$0.052) & 0.604($\pm$0.141) & 0.468($\pm$0.095) & 0.570($\pm$0.078) & 0.429($\pm$0.039) & 0.439($\pm$0.067) & 0.663($\pm$0.058) & 0.468($\pm$0.095) & 0.577($\pm$0.081) & 0.62($\pm$0.014) & 0.576($\pm$0.02) & 0.547($\pm$0.015)& 0.729($\pm$0.102) &	0.682($\pm$0.13) \\ 
        \textbf{OPRK1} & 0.822($\pm$0.088) & 0.782($\pm$0.106) & 0.811($\pm$0.099) & 0.852($\pm$0.082) & 0.840($\pm$0.074) & 0.743($\pm$0.147) & 0.646($\pm$0.15) & 0.801($\pm$0.107) & 0.763($\pm$0.143) & 0.65($\pm$0.13) & 0.722($\pm$0.195) & 0.65($\pm$0.161) & 0.801($\pm$0.115) & 0.756($\pm$0.14) & 0.653($\pm$0.125) & 0.778($\pm$0.024) & 0.594($\pm$0.046) & 0.55($\pm$0.007) & 0.790($\pm$0.246) & 0.728($\pm$0.275)\\ 
        \textbf{PKM2} & 0.646($\pm$0.039) & 0.75($\pm$0.033) & 0.784($\pm$0.033) & 0.801($\pm$0.038) & 0.787($\pm$0.023) & 0.59($\pm$0.044) & 0.659($\pm$0.033) & 0.68($\pm$0.032) & 0.743($\pm$0.023) & 0.768($\pm$0.019) & 0.594($\pm$0.043) & 0.654($\pm$0.03) & 0.677($\pm$0.034) & 0.742($\pm$0.022) & 0.767($\pm$0.019) & 0.678($\pm$0.005) & 0.647($\pm$0.008) & 0.55($\pm$0.007)&0.813($\pm$0.036)& 0.786($\pm$0.042) \\ 
        \textbf{PPARG} & 0.602($\pm$0.101) & 0.691($\pm$0.096) & 0.834($\pm$0.037) & 0.778($\pm$0.062) & 0.797($\pm$0.052) & 0.643($\pm$0.072) & 0.559($\pm$0.114) & 0.577($\pm$0.222) & 0.591($\pm$0.25) & 0.504($\pm$0.259) & 0.559($\pm$0.144) & 0.561($\pm$0.11) & 0.571($\pm$0.22) & 0.594($\pm$0.251) & 0.495($\pm$0.257) & 0.654($\pm$0.023) & 0.701($\pm$0.019) & 0.692($\pm$0.018) & 0.837($\pm$0.209)& 0.728($\pm$0.211)\\ 
        \textbf{TP53} & 0.581($\pm$0.133) & 0.69($\pm$0.074) & 0.738($\pm$0.055) & 0.724($\pm$0.092) & 0.771($\pm$0.063) & 0.738($\pm$0.103) & 0.705($\pm$0.071) & 0.821($\pm$0.035) & 0.795($\pm$0.055) & 0.842($\pm$0.057) & 0.745($\pm$0.108) & 0.709($\pm$0.064) & 0.821($\pm$0.04) & 0.795($\pm$0.055) & 0.844($\pm$0.056) & 0.609($\pm$0.026) & 0.536($\pm$0.02) & 0.503($\pm$0.018) &0.717($\pm$0.144)& 0.712($\pm$0.116)\\ 
        \textbf{VDR} & 0.649($\pm$0.037) & 0.712($\pm$0.026) & 0.754($\pm$0.021) & 0.817($\pm$0.032)	& 0.833($\pm$0.015) & 0.578($\pm$0.028) & 0.652($\pm$0.03) & 0.717($\pm$0.033) & 0.786($\pm$0.025) & 0.809($\pm$0.027) & 0.568($\pm$0.03) & 0.644($\pm$0.036) & 0.716($\pm$0.034) & 0.786($\pm$0.025) & 0.811($\pm$0.026) & 0.747($\pm$0.009) & 0.674($\pm$0.009) & 0.634($\pm$0.007)&0.888($\pm$0.027)&0.829($\pm$0.028)\\ \hline
        \textbf{COVID-19} & 0.608($\pm$0.142) & 0.692($\pm$0.077) & 0.745($\pm$0.084) & 0.771($\pm$0.065)& n/a & 0.451($\pm$0.094) & 0.687($\pm$0.067) & 0.888($\pm$0.028) & 0.901($\pm$0.032) & n/a & 0.515($\pm$0.091) & 0.598($\pm$0.098) & 0.886($\pm$0.029) & 0.901($\pm$0.029) & n/a & 0.651($\pm$0.015) & 0.447($\pm$0.027) & 0.632($\pm$0.017) & 0.713($\pm$0.137) & 0.590($\pm$0.049)\\
    \hline \\ [0.5ex]
    \end{tabular}%
\end{adjustbox}
\label{table:PCA results}
\end{sidewaystable}

\begin{figure}[!]
    \centering
    \includegraphics[width=0.75\textwidth]{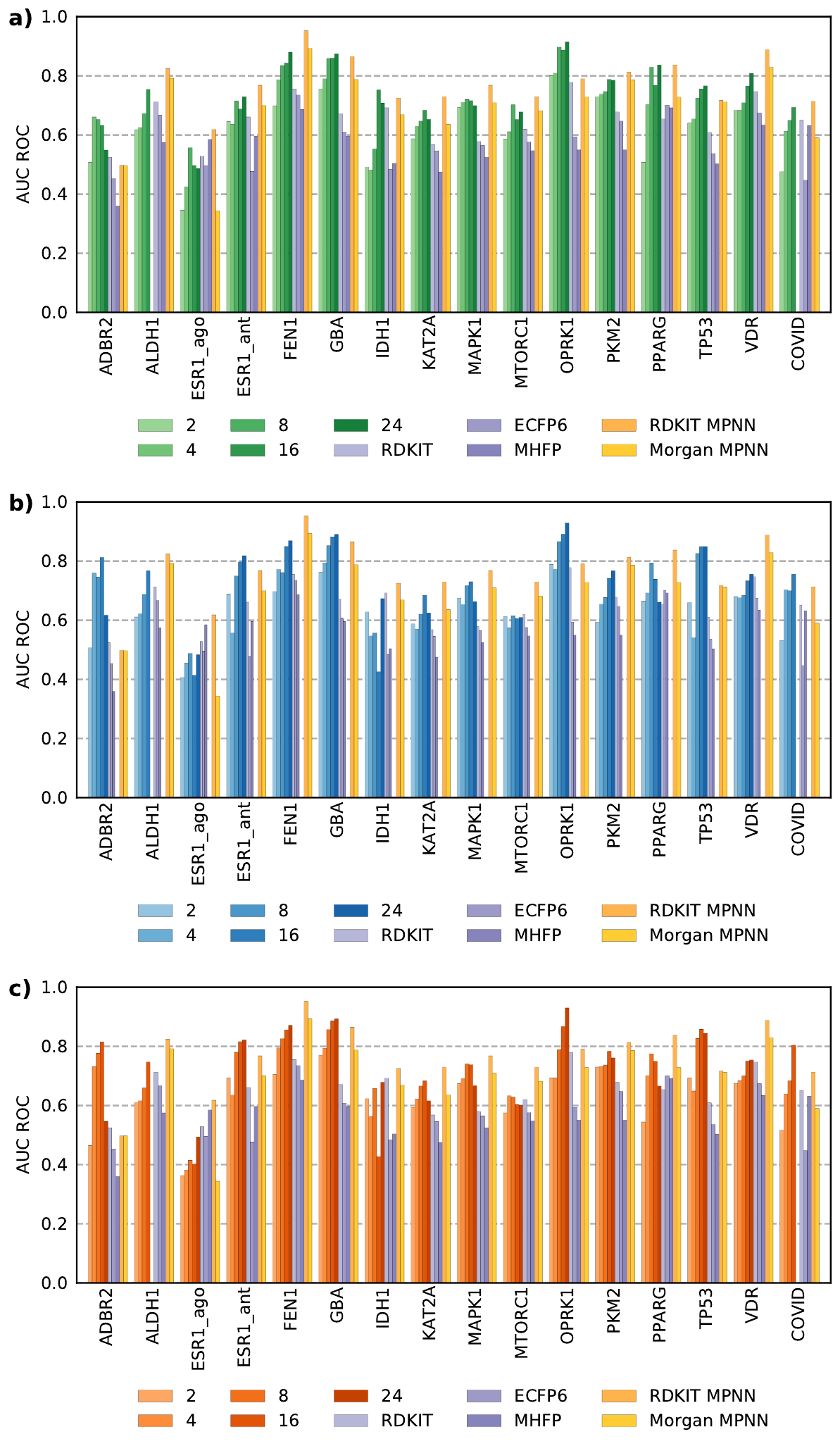}
    \caption{Graphical representation of mean AUC ROC results (standard deviation omitted for ease of representation) for benchmark LIT-PBCA and COVID-19 dataset from Table~\ref{table:ANOVA results} using \textbf{ANOVA} features selection. \textbf{a)} Mean CSVC AUC ROC, per feature (green) and against RMD (purple) and MPNN (yellow) results. \textbf{b)} Mean QSVC tuned with CSVC $C$ parameter (blue). \textbf{c)}  Mean QSVC tuned with default $C$ parameter (red). }
    \label{histoAppendixANOVA}
\end{figure}

\begin{figure}
    \centering
    \includegraphics[width=0.75\textwidth]{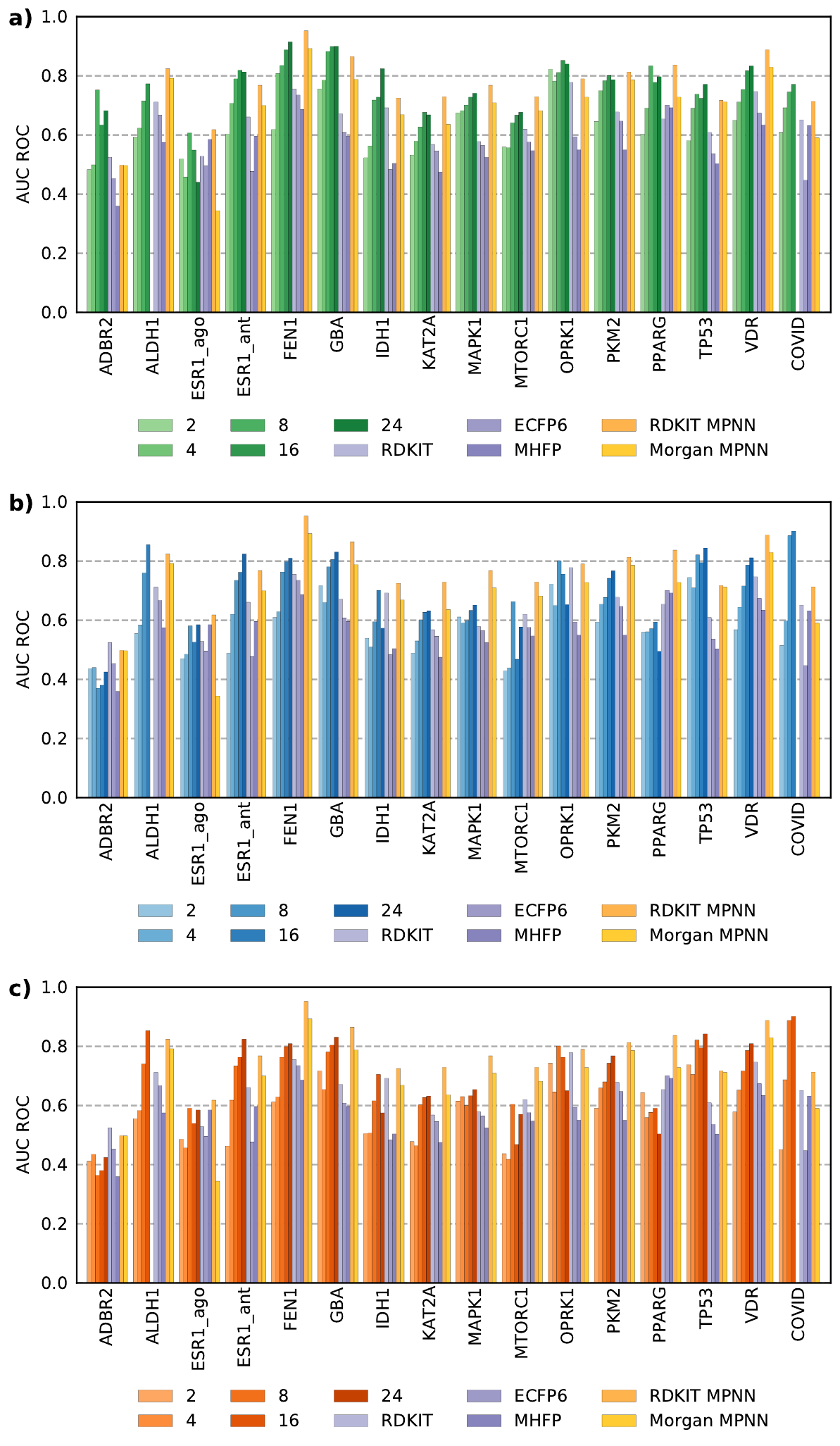}
    \caption{Graphical representation of mean AUC ROC results (standard deviation omitted for ease of representation) for benchmark LIT-PBCA and COVID-19 datasets from Table~\ref{table:PCA results}, using \textbf{PCA} features selection. \textbf{a)} Mean CSVC AUC ROC, per feature (green) and against RMD (purple) and MPNN (yellow) results. \textbf{b)} Mean QSVC tuned with CSVC $C$ parameter (blue). \textbf{c)}  Mean QSVC tuned with default $C$ parameter (red). }
  \label{histoAppendixPCA}
\end{figure}


\end{document}